\documentclass[useAMS,usenatbib]{mn2e}
\usepackage{graphicx}

\title[Testing the `dark-energy'-dominated cosmology via the Solar-system
experiments]%
{Testing the `dark-energy'-dominated cosmology via the Solar-system
experiments}
\author[Yu. V. Dumin]{Yu. V. Dumin%
\thanks{E-mail: dumin@yahoo.com}\\
Theoretical Department, IZMIRAN, Russian Academy of Sciences,
Troitsk, Moscow region 142190, Russia%
\thanks{Permanent address for correspondence.}\\
IPAM Building, University of California, Los Angeles,
460 Portola Plaza, CA 90095-7121, USA}

\begin{document}

\date{Accepted Year Month Day. Received 2006 October 6;
      in original form 2006 March 6}

\pagerange{\pageref{firstpage}--\pageref{lastpage}} \pubyear{2006}

\maketitle

\label{firstpage}

\begin{abstract}
The effect of `dark energy' (i.e.\ the $\Lambda$-term in Einstein
equations) is sought for at the interplanetary scales by comparing
the rates of secular increase in the lunar orbit obtained by two
different ways:
(1)~measured immediately by the laser ranging and
(2)~estimated independently from the deceleration of the Earth's proper
rotation.
The first quantity involves both the well-known effect of
geophysical tides and the Kottler effect of $\Lambda$-term
(i.e.\ a kind of the `local' Hubble expansion), while
the second quantity is associated only with the tidal influence.
The difference between them, 2.2$\pm$0.3~cm{\,}yr${}^{-1}$,
can be attributed just to the local Hubble expansion with rate
$H_0^{\rm (loc)}\! = 56{\pm}8$~km{\,}s${}^{-1}${\,}Mpc${}^{-1}$.
Assuming that Hubble expansion is formed locally only by
the uniformly distributed dark energy ($\Lambda$-term),
while globally also by a clumped substance (for the most part,
the cold dark matter), the total (large-scale) Hubble constant
should be $H_0 = 65{\pm}9$~km{\,}s${}^{-1}${\,}Mpc${}^{-1}$.
This is in reasonable agreement both with the commonly-accepted
WMAP result, $H_0 = 71{\pm}3.5$~km{\,}s${}^{-1}${\,}Mpc${}^{-1}$,
and with the data on supernovae Ia distribution.
The above coincidence can serve as one more argument in favor of
the dark energy.
\end{abstract}


\section{Introduction}

According to the recent astronomical data, the most part of energy
density in the Universe (up to 75\%) is in the `dark' form (such as
the so-called `quintessence', inflaton potential, polarization of
vacuum and so on), which is effectively described by $\Lambda$-term in
the Einstein equations \citep[e.g.\ reviews by][etc.]{vdb99,che01,kra04}.
All arguments in favor of the dark energy were obtained so far
from the observational data related to very large (intergalactic)
scales, such as a distribution of supernovae Ia as function of their
redshift, the spectrum of fluctuations of the cosmic microwave
background radiation, the spectra of absorption lines from distant
sources, or Ly$\alpha$ forest, and so on.

Is it possible to find a manifestation of the dark energy at
much less scales (e.g.\ inside the Solar system)?
In general, such effects can be expected from the solution of
the equations of General Relativity for a point-like mass~$M$ in
the $\Lambda$-dominated (de~Sitter) Universe, which was obtained
by~\citet{kot18} a very short time after the original Schwarzschild
solution. (More details are given in Appendix~\ref{sec:KottMet}.)

The presence of $\Lambda$-term should change, particularly,
the standard relativistic shift of Mercury's perihelion,
predicted by General Relativity.
This was the idea by~\citet{car98}, who proposed using
the measure of the uncertainty in our knowledge of Mercury's
perihelion shift to impose the upper bound on~$\Lambda$.
The result obtained was not so good as other cosmological estimates
but, surprisingly, the accuracy was worse by only $1{\div}2$~orders
of magnitude. So, according to the above-cited authors,
improvement of the value of the shift by one or two decimal digits
should make such method of determination of~$\Lambda$ competitive
with the observations at large scales.
A more skeptical viewpoint on the same subject was presented
recently by \citet{ior06}.

In any case, since accuracy of the above method is still insufficient,
it was proposed in our previous papers \citep{dum01,dum03}
to utilize the data of radial (rather than angular) measurements of
the Moon to reveal anomalous increase in its orbit produced by the
$\Lambda$-term in metric~(\ref{eq:metric_cosm_coord})--(\ref{eq:g_angl}),
under assumption that the central mass~$M$ belongs to the Earth.
This looks formally as `local' Hubble expansion. Unfortunately,
the result of the earlier works was quite strange:
the `local'  Hubble constant was found to be approximately in the middle
between zero and the standard intergalactic value,
which did not allow a reasonable quantitative interpretation.

The aim of the present work is to describe a much improved analysis
of the available observations and to show that its results are in
reasonable theoretical agreement with the large-scale data.

\section{Theoretical viewpoints on the problem of local Hubble expansion}

The secular increase in planetary radii due to the $\Lambda$-term,
discussed in Appendix~\ref{sec:KottMet}, formally looks like a kind of
the Hubble expansion. So, let us briefly discuss why is it necessary
to reexamine the problem of local Hubble expansion just in the context
of `dark-energy'-dominated cosmological models?

In general, Hubble dynamics at the small scales is studied for a long
time, starting from the pioneering work by~\citet{mvi33}. Although
the results by various authors obtained by now were quite contradictory
\citep[e.g.\ review by][and references therein]{bon00}, the most popular
point of view was that the Hubble expansion manifests itself only at
the sufficiently large distances (from a few Mpc) and is absent at
the less scales at all \citep[e.g.][]{mis73}. There were a few arguments in
favor of such conclusion.

\begin{figure}
\begin{center}
\vspace*{8pt}
\includegraphics[width=6.8cm]{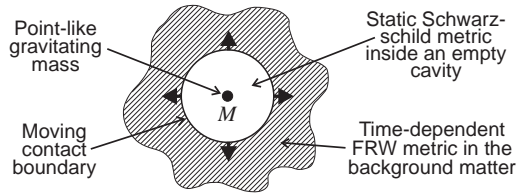}
\end{center}
\caption{Schematic illustration of the Einstein--Straus theorem.
\label{fig:ES_Theorem}}
\end{figure}

The first of them is based on the so-called Einstein--Straus
theorem \citep{ein45}. Let us consider the Friedmann--Robertson--Walker
(FRW) cosmological model, uniformly filled with some kind of matter.
Next, let us assume that the background matter inside a specified sphere
is cut off and concentrated in the point in its centre, as illustrated
in Fig.~\ref{fig:ES_Theorem}. Then, using equations of General Relativity,
it can be shown that gravitational field inside the empty cavity
is described by the purely static Schwarzschild metric and
begins to experience the cosmological expansion only outside
its contact boundary with the background matter distribution;
this boundary moving just with the Hubble velocity.
Therefore, there is no any Hubble expansion in the vicinity of
the mass~$M$.

Unfortunately, despite a mathematical elegance of this result,
it is absolutely unclear how can the above-mentioned cavities be
identified for the real astronomical objects. Moreover, this theorem
becomes evidently inapplicable at all in the case of
`dark-energy'-dominated cosmology, because it is meaningless
to consider an empty cavity in the vacuum energy distribution.

The second argument against the Hubble expansion at small scales
is based on a quasi-Newtonian treatment of Hubble effect
in a small volume as a tidal-like action by distant matter
\citep[e.g.\ the recent work by][and references therein]{dom01}.
The final conclusion usually derived by this way is that
there should be no Hubble expansion in the gravitationally-bound
systems (i.e.\ the ones whose kinetic energy is less than potential
energy), such as the planetary systems or stars in galaxies.
According to this criterion, the Hubble expansion should manifest itself
only from $5{\div}10$~Mpc, as was actually observed for a long time.

Unfortunately, the tidal treatment of Hubble effect is not
well justified from the theoretical point of view.
Besides, more accurate astronomical measurements in the recent
few years revealed a well-formed Hubble flow down to the distances of
about 1~Mpc, which are considerably less than the scale of
gravitational unbinding \citep{che01,ekh01}.
Moreover, in the $\Lambda$-dominated cosmology the tidal effects
cannot be of primary significance just because of the perfectly
uniform distribution of the dark energy and, therefore,
may be expected only from minor constituents of the Universe.

At last, one more approach for treating the influence of
cosmic expansion on the dynamics of small-scale systems is based
on the Einstein--Infeld--Hoffmann (EIH) surface integral method,
which enables to derive the equations of motion of the particles
immediately from the field equations. This method is based on
the integration of field equations over the small closed surfaces
surrounding the point-like sources of the field and subsequent
use of independence of these integrals on the particular surfaces.
Such approach is widely employed to obtain the post-Newtonian equations
of motion against the background of flat (Minkowski) space--time
and was applied also to the problem of local Hubble expansion by
\citet{and95}. It was found that planetary systems should really
expand but with a rate much less than for the entire Universe.
Unfortunately, the EIH method also becomes inappropriate if
the space is filled everywhere with the dark energy, because
the surface integrals are no longer invariant when
the integration surfaces are moved.

Finally, a frequent experimental argument against Hubble expansion
within the Solar system is based on the available constraint
on time variation in the gravitational constant derived from
the planetary dynamics, which is now as strong as
$ \, {\dot G} / G = \! (4{\pm}9){\times}10^{-13} $~yr${}^{-1}$
\citep{wil04}. So, as was concluded by these authors,
``the $ {\dot G} / G $ uncertainty is 83~times smaller than
the inverse age of the Universe, $ t_0 = 13.4 $~Gyr~\dots \,
Any isotropic expansion of the Earth's orbit which conserves
angular momentum will mimic the effect of $ \dot G $
on the Earth's semimajor axis,
$ {\dot a} / a = - {\dot G} / G $~\dots \,
There is no evidence for such local (${\sim}1$~AU) scale expansion
of the solar system.''

\begin{table*}
\begin{center}
\begin{minipage}{118mm}
\caption{Rates of secular increase in the mean Earth--Moon distance.
\label{tab:Compar}}
\begin{tabular}{lll}
\hline
\multicolumn{1}{c}{Method} &
\multicolumn{1}{c}{Immediate measurement by} &
\multicolumn{1}{c}{Independent estimate from the}
\\
& \multicolumn{1}{c}{the lunar laser ranging%
\footnote{The data were taken from the review by \citet{dic94}.}} &
\multicolumn{1}{c}{Earth's tidal deceleration%
\footnote{
Formula~(\ref{eq:tidal_effect}) was used with the rate of Earth's
diurnal deceleration~(\ref{eq:T_E}).}}
\\ \hline
Effects involved &
(1) geophysical tides &
(1) geophysical tides \\
&
(2) local Hubble expansion &
\\
Numerical value &
$3.8{\pm}0.1$~cm{\,}yr${}^{-1}$ &
$1.6{\pm}0.2$~cm{\,}yr${}^{-1}$
\\ \hline
\end{tabular}
\end{minipage}
\end{center}
\end{table*}

Unfortunately, the above-stated equivalence between the effect of
variable $G$ and the cosmological expansion is based solely on
the Newtonian arguments. A more accurate treatment of this problem
in the framework of General Relativity for a general case of
the multi-component Universe is very difficult. Nevertheless,
it can be performed in analytic form for the particular case of
a point-like mass in the Universe filled only with dark energy
(the $\Lambda$-term), and the corresponding results are outlined
in Appendix~\ref{sec:KottMet}. As follows from
expressions~(\ref{eq:g_tt_approx})--(\ref{eq:g_angl_approx}),
the $\Lambda$-dependence of a few terms of the resulting
metric tensor can really be reinterpreted as the effect of
variable $G$, but this is not true in general:
there are some terms whose $\Lambda$-dependence is
radically different from any variations in $G$.

\section{Analysis of observational data}

Since all the commonly-used arguments against the small-scale Hubble
expansion fail in the case of dark energy, it becomes reasonable
to seek for the corresponding effect; and the most sensitive tool
seems to be the lunar laser ranging (LLR) \citep[e.g.][]{dic94,nor99}.
For example, if we assume that planetary systems experience
the Hubble expansion with the same rate as everywhere in the Universe
($60{\div}70$~km{\,}s${}^{-1}${\,}Mpc${}^{-1}$), then average radius
of the lunar orbit $R$ should increase by approximately 50~cm for
the period of 20~years. On the other hand, the accuracy of
LLR measurements during the last 20~years was maintained at
the level of $2{\div}3$~cm \citep{dic94}; so the perspective of
revealing the local Hubble effect looks very good.

The main obstacle that needs to be got around is to exclude
the effect of geophysical tides, which also contributes to
the secular increase in the Earth--Moon distance.
As is known \citep[e.g.][]{kau68}, because of dissipative effects,
the Earth's tidal bulge, formed by the lunar attraction,
is not perfectly aligned in the direction to the Moon but
slightly shifted towards the Earth's proper rotation.
Therefore, there is a torque moment, which decreases
the proper angular momentum of the Earth and increases
the orbital momentum of the Moon. As a result, the average
Earth--Moon distance~$R$ gradually increases in the course of time.

According to the law of conservation of angular momentum,
the secular variation in $R$ is related to the change in
the Earth's diurnal period $T_{\rm E}$ by the simple formula:
\begin{equation}
\dot R = k \, {\dot T}_{\rm E} \, ,
\label{eq:tidal_effect}
\end{equation}
where $k \! = \! 1.81{\times}10^5$~cm{\,}s${}^{-1}$
\citep[a more detailed discussion can be found, for example,
in our previous work,][]{dum03}.
So, if ${\dot T}_{\rm E}$ is known from independent
astrometric measurements of the Earth's rotation deceleration
with respect to distant objects, then relation~(\ref{eq:tidal_effect})
can be used to exclude the influence of geophysical tides and,
thereby, to reveal a probable presence of local Hubble expansion.

Unfortunately, ${\dot T}_{\rm E}$ is not a well-defined quantity:
its values derived from the observations in telescopic era and
from the various sets of pre-telescopic data appreciably differ
from each other \citep[e.g.][]{ste84}. It is not so important for
us now what is the reason for these discrepancies: this may be
the systematic errors in the earliest data or, for example,
the tectonic processes appreciably changed the Earth's moment of
inertia just before the period of telescopic observations.
Since LLR data refer to the last decades, they should be confronted
with the most recent values of the Earth's rotation deceleration.
The corresponding telescopic data, starting from the middle of
the 17th century, were processed by a few researches; and one of
the most detailed compilations was presented recently by \citet{sid02}.

Of course, the value of secular trend derived from the quite short
time series can suffer from the considerable periodic and
quasi-periodic variations in $T_{\rm E}$. So, the main aim of
our statistical analysis outlined in Appendix~\ref{sec:StatAn}
was to estimate as carefully as possible the `mimic' effect of
such variations. Taking into account the corresponding uncertainty,
the resulting value can be written as
\begin{equation}
{\dot T}_{\rm E} \! =
  \! (8.77{\pm}1.04){\times}10^{-6}~{\rm s}{\,}{\rm yr}^{-1} .
\label{eq:T_E}
\end{equation}

The results of the entire analysis of LLR vs.\ the astrometric data
are summarized in Table~\ref{tab:Compar}. The excessive rate of increase
of the lunar orbit $2.2{\pm}0.3$~cm{\,}yr${}^{-1}$ can be attributed to
the local Hubble expansion with rate
\begin{equation}
H_0^{\rm (loc)} = 56{\pm}8~{\rm km{\,}s^{-1}{\,}Mpc^{-1}} .
\label{eq:H_0^loc}
\end{equation}
This value is appreciably greater than in our earlier work \citep{dum03},
where it was found to be only $33{\pm}5$~km{\,}s${}^{-1}${\,}Mpc${}^{-1}$.
This was because of using a substantially different rate of the Earth's
deceleration,
${\dot T}_{\rm E} \! = \! 1.4{\times}10^{-5}$~s{\,}yr${}^{-1}$.
The last-mentioned value was obtained for the first time by
\citet{ste84}, who used the telescopic observations
supplemented by a much longer series of medieval Arabian data;
and their result was inaccurately cited in a number of subsequent
reviews and monographs \citep[e.g.][]{per00} as derived from
the telescopic observations alone.

Finally, it should be mentioned that the basic
relation~(\ref{eq:tidal_effect}) was written under assumption of
constant moment of inertia of the Earth, which is a questionable item.
For example, over 20~years ago \citet{yod83} found that Earth's
oblateness, commonly characterized by the gravitational
harmonic coefficient $J_2$, was decreasing. This was interpreted as
viscous rebound of the solid Earth from the decrease in load due to
the last deglaciation. The observed secular effect
$ {\dot J}_2 \! = \! -3{\times}10^{-11}$~yr${}^{-1}$ resulted in
the Earth's spin-up due to decreasing moment of inertia and,
thereby, enabled the above-cited authors to get a reasonable
agreement between the various sets of data on the Earth's rotation.
Therefore, following this approach, it would be unnecessary
to take into consideration any other influences,
such as the cosmological Hubble expansion.

Unfortunately, the early results by \citet{yod83} were not
confirmed by the most recent studies. For example, as follows from
the analysis by \citet{bou04} performed over a sufficiently long
time interval (1985--2002), the coefficient~$J_2$ has
a much smaller secular trend but a considerable oscillatory component
with a period of two decades.
An even more striking disagreement with the earlier data
was obtained by \citet{cox02}, who found that since 1997 or 1998
the secular trend in~$J_2$ has approximately the same absolute value
as reported by \citet{yod83} but the opposite sign
(namely, $ {\dot J}_2 \! = \! +2.2{\times}10^{-11}$~yr${}^{-1}$).
Therefore, we can conclude that
(1)~a considerable disagreement between the LLR and astrometric data
still exists and
(2)~the coefficient~$J_2$ experiences most probably
a quasi-periodic variation with a typical time scale of a few decades.
The last-mentioned property justifies usage of
formula~(\ref{eq:tidal_effect}), because the temporal variations
in the Earth's moment of inertia should be averaged out
when data on the Earth's rotation are taken for the period of
${\sim}350$~years.

\section{Theoretical interpretation}
\label{sec:Interp}

How can the value~(\ref{eq:H_0^loc}) be interpreted?
It is reasonable to assume that local Hubble expansion is formed
only by the uniformly-distributed dark energy ($\Lambda$-term),
while the irregularly-distributed (aggregated) forms of matter begin
to affect the Hubble flow at the larger distances,
thereby increasing its rate up to the standard intergalactic value.

If the universe is spatially flat and filled only with vacuum and
a dust-like (`cold') matter, with densities $\rho_{{\Lambda}0}$ and
$\rho_{{\rm D}0}$ respectively, then
\begin{equation}
H_0 = \, \sqrt{\frac{8 \pi G}{3}}
      \; \sqrt{\, \rho_{{\Lambda}0} + \rho_{{\rm D}0}}
\label{eq:H_0-rho}
\end{equation}
\citep[e.g.][]{lan75}.
So, if $H_0$ is formed locally only by ${\rho}_{{\Lambda}0}$,
while globally by both these terms, ${\rho}_{{\Lambda}0}$ and
${\rho}_{{\rm D}0}$ (or, in terms of the relative densities,
$ {\Omega}_{{\Lambda}0} = {\rho}_{{\Lambda}0} / {\rho}_{\rm cr} $
and $ {\Omega}_{{\rm D}0} = {\rho}_{{\rm D}0} / {\rho}_{\rm cr} $),
then
\begin{equation}
\frac{H_0^{\rm (loc)}}{H_0} =
{\left[ 1 +
  \frac{\Omega_{{\rm D}0}}{\Omega_{{\Lambda}0}} \, \right]}^{-1/2} .
\end{equation}

At the commonly-accepted values $\Omega_{{\Lambda}0} \! = \! 0.75$
and $\Omega_{{\rm D}0} \! = \! 0.25$, we get
${H_0} / {H_0^{\rm (loc)}} \! \approx 1.15$. Therefore,
\begin{equation}
H_0 = 65{\pm}9~{\rm km{\,}s^{-1}{\,}Mpc^{-1}} \, ,
\label{eq:H_0}
\end{equation}
which is in reasonable agreement both with the well-known WMAP result,
$71{\pm}3.5$~km{\,}s${}^{-1}${\,}Mpc${}^{-1}$, and with the most recent
Hubble diagram for a complete sample of type Ia supernovae \citep{rei05},
whose interpretation requires a slightly reduced value of $H_0$.

On the other hand, at the given ratio ${H_0^{\rm (loc)}}\!/{H_0}$
we have
\begin{equation}
\frac{\Omega_{{\rm D}0}}{\Omega_{{\Lambda}0}} =
  \, {\bigg( \frac{H_0}{H_0^{\rm (loc)}} \bigg)}^{\!\! 2}
  \, - 1 \; .
\label{eq:Omega_ratio}
\end{equation}
So, using the most popular value of the total Hubble constant
$H_0 \! = \! 71$~km{\,}s${}^{-1}${\,}Mpc${}^{-1}$
and our value of the local Hubble constant
$H_0^{\rm (loc)} \! = \! 56$~km{\,}s${}^{-1}${\,}Mpc${}^{-1}$, we get
${\Omega_{{\rm D}0}}/{\Omega_{{\Lambda}0}} \approx 0.6$, i.e.\ a much
less fraction of the dark energy than the commonly-accepted one.
Therefore, a slightly reduced value of $H_0$ seems to be a preferable
option.

\section{Conclusions}

As follows from the above analysis, the presence $\Lambda$-term
can give us a reasonable explanation of the anomalous increase
in the lunar orbit, consistent with the large-scale astronomical
data. Thereby, this is one more argument in favor of the dark energy.

Besides, if the local Hubble expansion really exists, it should result
in profound consequences not only for cosmological evolution but
also for the dynamics of planetary systems and other `small-scale'
astronomical phenomena, which have to be studied in more detail.

\section*{Acknowledgments}

I am grateful to
Yu.V.~Baryshev,
P.L.~Bender,
V.~Doku\-chaev,
M.~Fil'chenkov,
S.S.~Gershtein,
C.~Horellou,
I.B.~Khrip\-lovich,
B.V.~Komberg,
C.~Laemmerzahl,
S.M.~Mo\-lodensky,
J.~Mueller,
T.~Murphy,
P.J.E.~Peebles,
A.~Poludnenko,
A.I.~Rez,
A.~Ruzmaikin,
M.~Sereno,
A.~Starobinsky,
N.I.~Sha\-kura,
G.~Tammann,
A.V.~Toporensky,
S.~van den Bergh
and the unknown referee
for valuable discussions and critical comments.
I am especially grateful to K.S.J.~Anderson (Apache Point Observatory)
for careful checking the formulas and pointing to a misprint.

This work was partially performed in the framework of
Grand Challenge Problems in Computational Astrophysics Program,
headed by M.~Morris (UCLA) and funded by the National
Science Foundation.

\appendix

\section{A point-like mass in the Lambda-dominated Universe}
\label{sec:KottMet}

Solution of the equations of General Relativity for a point-like
mass~$M$ in the Universe filled only with $\Lambda$-term is
\begin{eqnarray}
{\rm d} s^2 \!\! & \!\!\! = \!\! & \! -\, {\Bigl( 1 - \frac{2 G M}{c^2 r'}
  - \frac{\Lambda {r'}^2}{3} \Bigr)}\, c^2 {\rm d}{t'}^2
\nonumber
\\
  & & \! + \, {\Bigl( 1 - \frac{2 G M}{c^2 r'}
  - \frac{\Lambda {r'}^2}{3} \Bigr)}^{\!\! -1} \! {\rm d}{r'}^2 \!
\nonumber
\\
  & & \! + \:\: {r'}^2 ( {\rm d}{\theta}^2 \!
  + {\sin}^2{\theta} \, {\rm d}{\varphi}^2 )
\label{eq:Kottler_metric}
\end{eqnarray}
\citep{kot18}, where $G$ is the gravitational constant,
and $c$ is the speed of light
\citep[for general review, see also][]{kra80}.

After a transformation to the cosmological Robertson--Walker coordinates
\begin{equation}
r' = \, a_0 \, {\exp} \Bigl( \frac{\displaystyle c t}%
     {\displaystyle r_0} \Bigr) \, r \, ,
\label{eq:r_coord}
\end{equation}
\begin{equation}
t' = \, t - \frac{1}{2} \, \frac{\displaystyle r_0}{\displaystyle c} \:\:
     {\ln} \Bigl[ 1 - \frac{\displaystyle a_0^2}{\displaystyle r_0^2} \:
     {\exp} \Bigl( \frac{\displaystyle 2 c t}{\displaystyle r_0} \Bigr)
     \: r^2 \Bigr] \, ,
\label{eq:t_coord}
\end{equation}
metric~(\ref{eq:Kottler_metric}) takes the form
\begin{eqnarray}
{\rm d} s^2 \!\!\! & \!\!\! = \!\! & \!\!\! g_{tt} \, c^2 {\rm d}{t}^2
  + \, 2 \, g_{tr} \, c \, {\rm d}{t} \, {\rm d}{r}
  + \, g_{rr} \, {\rm d}{r}^2
\nonumber
\\
  & \! + & \!\!\! g_{\theta \theta} \, {\rm d}{\theta}^2
  + \,g_{\varphi \varphi} \, {\rm d}{\varphi}^2 \, ,
\label{eq:metric_cosm_coord}
\end{eqnarray}
where
\begin{equation}
g_{tt} =
  \frac{\displaystyle - {\Bigl( 1 - \frac{r_g}{r'} - \frac{{r'}^2}{r_0^2}
\Bigr)}^{\!\! 2} + {\Bigl( 1 - \frac{{r'}^2}{r_0^2} \Bigr)}^{\!\! 2}
\frac{{r'}^2}{r_0^2}}%
       {\displaystyle {\Bigl( 1 - \frac{r_g}{r'} - \frac{{r'}^2}{r_0^2}
\Bigr)} {\Bigl( 1 - \frac{{r'}^2}{r_0^2} \Bigr)}^{\!\! 2} } \:\: ,
\label{eq:g_tt}
\end{equation}

\begin{equation}
g_{tr} =
  \frac{\displaystyle {\Bigl( 1 - \frac{{r'}^2}{r_0^2} \Bigr)}^{\!\! 2}
- {\Bigl( 1 - \frac{r_g}{r'} - \frac{{r'}^2}{r_0^2} \Bigr)}^{\!\! 2}}%
       {\displaystyle {\Bigl( 1 - \frac{r_g}{r'} - \frac{{r'}^2}{r_0^2}
\Bigr)} {\Bigl( 1 - \frac{{r'}^2}{r_0^2} \Bigr)}^{\!\! 2}}
  \; \frac{r'}{r_0} \, \frac{r'}{r} \:\: ,
\label{eq:g_tr}
\end{equation}

\begin{equation}
g_{rr} =
  \frac{\displaystyle {\Bigl( 1 - \frac{{r'}^2}{r_0^2} \Bigr)}^{\!\! 2}
- {\Bigl( 1 - \frac{r_g}{r'} - \frac{{r'}^2}{r_0^2} \Bigr)}^{\!\! 2}
\frac{{r'}^2}{r_0^2} }%
       {\displaystyle {\Bigl( 1 - \frac{r_g}{r'} - \frac{{r'}^2}{r_0^2}
\Bigr)} {\Bigl( 1 - \frac{{r'}^2}{r_0^2} \Bigr)}^{\!\! 2}}
  \; \frac{{r'}^2}{r^2} \:\: ,
\label{eq:g_rr}
\end{equation}

\begin{equation}
\label{eq:g_angl}
g_{\theta \theta} = \, g_{\varphi \varphi} / {\sin}^2 {\theta} = {r'}^2 .
\end{equation}
In the above formulas,
$ r_g \! = 2 G M / c^2 $, $ r_0 = \sqrt{ 3 / \Lambda } \, $, and
$a_0$~is the scale factor of FRW Universe.

Taking $ a_0 \! = 1 $ at $ t \! = 0 $ and keeping only
the lowest-order terms of $ \, r_g $ and $ 1 / r_0 $, we get
\begin{equation}
g_{tt} \! \approx
  - \Bigl[ 1 - \frac{2 G M}{c^2 r}
  \Bigl( 1 - \frac{c \sqrt{\Lambda} \, t}{\sqrt{3}} \Bigr) \Bigr] \, ,
\label{eq:g_tt_approx}
\end{equation}

\begin{equation}
g_{tr} \! \approx \frac{4 \, G M \sqrt{\Lambda}}{\sqrt{3} \, c^2} \:\: ,
\label{eq:g_tr_approx}
\end{equation}

\begin{equation}
g_{rr} \! \approx
  \Bigl[ 1 + \frac{2 G M}{c^2 r}
  \Bigl( 1 - \frac{c \sqrt{\Lambda} \, t}{\sqrt{3}} \Bigr) \Bigr]
  \Bigl( 1 + \frac{2 c \sqrt{\Lambda} \, t}{\sqrt{3}} \Bigr) \, ,
\label{eq:g_rr_approx}
\end{equation}

\begin{equation}
g_{\theta \theta} = \, g_{\varphi \varphi} / {\sin}^2 {\theta} \approx
  \, r^2 \Bigl( 1 + \frac{2 c \sqrt{\Lambda} \, t}{\sqrt{3}} \Bigr) \, .
\label{eq:g_angl_approx}
\end{equation}

As is seen from the above expressions, manifestation of
the $\Lambda$-term in some components of the metric tensor
looks like the influence of variable $G$, if we assume that
$ G = G_0 + {\dot G} \, t $, where
$ {\dot G} = - c \sqrt{\Lambda} / \sqrt{3} \, $.
Unfortunately, such interpretation is not self-consistent:
the $\Lambda$-dependence of other components is irreducible
to the effect of variable $G$.

\section{Statistical analysis of the Earth's rotation deceleration}
\label{sec:StatAn}

It is well known that diurnal period of the Earth $T_{\rm E}$,
along with secular increase in the course of time, experiences
a wide range of oscillations with (quasi-)periods from
annual scale to many decades and, probably, even longer.
In fact, just the long-period variations enforced the most of
researchers to supplement the available telescopic data on
the Earth's rotation by ancient records of the solar eclipses
(whose reliability was, of course, not so good).

On the other hand, as was already mentioned in the main body of
the paper, we are interested in using only the data that are
as close as possible to the LLR measurements.
Of course, in deriving the secular term from the short data series,
a special care should be paid to a probable interference from
the oscillatory components: if periods of such oscillations are
comparable with the length of the series analyzed,
they can partially mimic the secular (linear) term and, therefore,
produce a substantial error in its final value.
So, the basic idea of our analysis was to simulate such mimic effect
and to find its maximum contribution to the secular term.

Let the observational data be fitted by the following function:
\begin{equation}
f(t) = f_0 + f_1 \, t +
       \, a \, \sin (2 \pi t / T) + \, b \, \cos (2 \pi t / T) \, ,
\label{eq:Regr_Fun}
\end{equation}
where $T$~is the trial period; and
$f_0$, $f_1$, $a$ and $b$~are the unknown regression
coefficients. They are determined, as usual, by minimization of
the functional
\begin{equation}
\Phi (f_0, f_1, a, b, T, N_{\rm b}, N_{\rm e}) =
  \sum_{i = N_{\rm b}}^{N_{\rm e}} {\left[ f(t_i) - f_i \, \right]}^2
\end{equation}
with respect to $f_0$, $f_1$, $a$ and $b$.
Here, $i$~is the number of year, $f_i$~is the corresponding
observed value of $f$; $N_{\rm b}$ and $N_{\rm e}$~are the beginning and
end of the analyzed time interval.

The resulting values of $f_0$, $f_1$, $a$ and $b$ will be, in general,
functions of $T$, $N_{\rm b}$ and $N_{\rm e}$. The extent of `mimic'
contribution from the periodic components to the secular term~$f_1$
(i.e.\ the contribution under the worst conditions) can be characterized
by the quantities
\begin{eqnarray}
f_1^{\rm (min)} (N_{\rm b}, N_{\rm e}) =
  \, \min_T f_1 (T, N_{\rm b}, N_{\rm e}) \, ,
\\
f_1^{\rm (max)} (N_{\rm b}, N_{\rm e}) =
  \, \max_T f_1 (T, N_{\rm b}, N_{\rm e}) \, ;
\end{eqnarray}
while $f_1^{\rm (lin)} (N_{\rm b}, N_{\rm e})$ will denote the value
derived with the purely linear regression function~(\ref{eq:Regr_Fun}),
when the coefficients $a$ and $b$ were dropped out \textit{a priopi}.

The second important task is to estimate the effect of truncation
of the time series on the resulting value of the secular term.
(Since the earliest telescopic data might be not so reliable,
it may be reasonable to exclude them from the analysis.)
So, we performed a number of calculations with
the shorter data series, when $N_{\rm b}$ changed from~1 to~50
(while $N_{\rm e}$ was always the same).

As was already mentioned, the primary observational data $f_i$ were
taken from the monograph by \citet{sid02}. They represent the mean
annual variations in the Earth's angular velocity
$ f(t) = ( \delta {\Omega}_{\rm E} / {\Omega}_{\rm E} )
\! \times \! 10^{10} $ for the years~1656 to~2000.

\begin{figure}
\begin{center}
\vspace*{6pt}
\includegraphics[width=7.4cm]{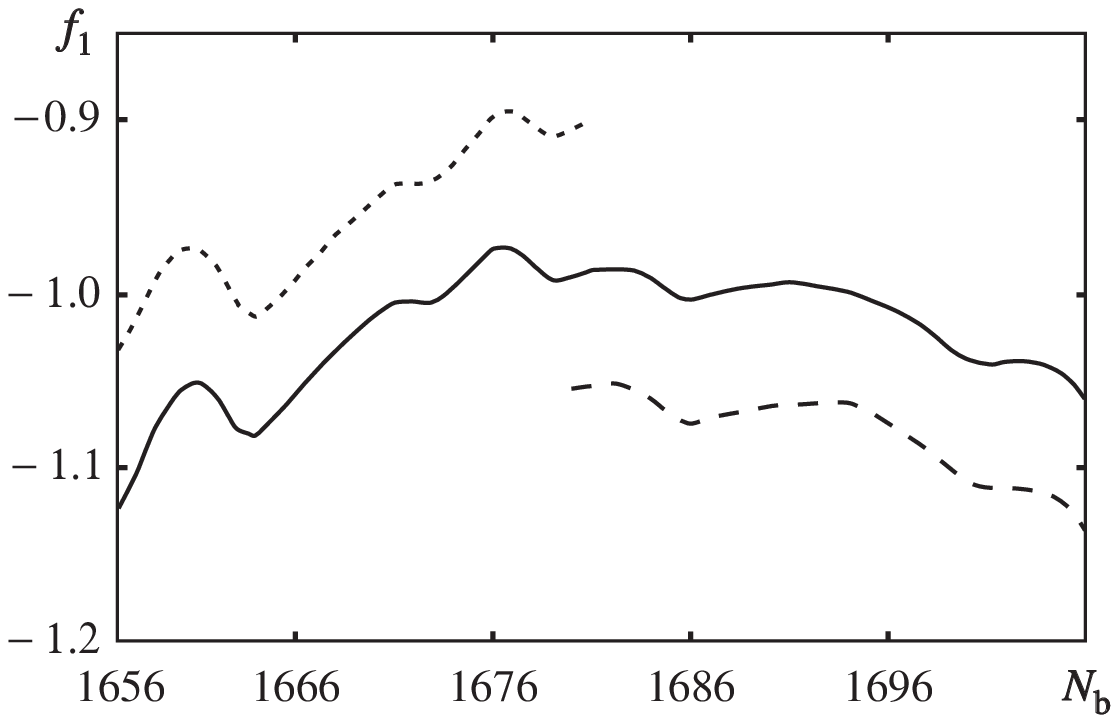}
\end{center}
\caption{Values of $f_1^{\rm (lin)}$~(solid curve),
$f_1^{\rm (min)}$~(dashed curve) and
$f_1^{\rm (max)}$~(dotted curve) as functions of
the beginning of the data series~$N_{\rm b}$.
In the intervals where some of the curves are not plotted,
the corresponding extrema were not found.
\label{fig:Regr_An}}
\end{figure}

The main results of our regression analysis are shown in
Fig.~\ref{fig:Regr_An}. As is seen, the secular term~$f_1$
can vary in total (as function of both $T$ and $N_{\rm b}$)
from $-1.135$~yr${}^{-1}$ to $-0.895$~yr${}^{-1}$.
So, its average value, distorted by the `mimic' effect of periodic
variations and insufficient accuracy of the earliest data,
can be written as $ f_1 = -1.015 \pm 0.12 $~yr${}^{-1}$.
Then, the required rate of the Earth's diurnal deceleration
will be $ {\dot T}_{\rm E} = - \, T_{\rm E} \, 10^{-10} f_1 $
(where $ T_{\rm E} = 8.64 \! \times \! 10^4 $~s), resulting in
${\dot T}_{\rm E} \! = \! (8.77{\pm}1.04){\times}10^{-6}$~s{\,}yr${}^{-1}$.

\bsp

\label{lastpage}

\end{document}